\documentclass[12pt]{article}
\usepackage{heck}
\usepackage{amsmath,amssymb}
\usepackage{hyperref}
\usepackage{graphicx}  

\graphicspath{{./Figures_orig/}}

\usepackage{xspace}
\newcommand*{\eg}{e.g.\@\xspace}
\newcommand*{\ie}{i.e.\@\xspace}
\makeatletter
\newcommand*{\etc}{%
    \@ifnextchar{.}%
        {etc}%
        {etc.\@\xspace}%
}
\makeatother

\def\i{\mathsf{i}}

\renewcommand{\d}{\mathrm{d}}
\newcommand{\br}{\overline}

\begin{document}


\title{New T-duality for Chern-Simons Theory}

\authors{\centerline{Masahito Yamazaki}}

\preprint{IPMU19-0049}

\date{April, 2019}

\institution{IPMU}{\centerline{Kavli IPMU (WPI), University of Tokyo, Kashiwa, Chiba 277-8583, Japan}}

\abstract{It has recently pointed out that a four-dimensional analog of Chern-Simons theory provides an elegant framework for understanding integrable models with spectral parameters. The goal of this short note is to better understand the relation of this theory to the more standard three-dimensional Chern-Simons theory. We point out that two Chern-Simons theories, in four dimensions and three dimensions, are related by a novel T-duality in field theory. We then discuss this T-duality in string theory. Our T-duality prescription applies to a more general class of topological quantum field theories, producing mixed topological/holomorphic theories. This paper is motivated by the observation by C.~Vafa.}


\maketitle 


\section{Introduction}

Recently, a new approach to integrable model has been proposed in
\cite{Costello:2013zra,Witten:2016spx,Costello:2017dso,Costello:2018gyb}. 

The starting point of this approach is the four-dimensional analog of the 
Chern-Simons theory \cite{Costello:2013zra}, whose action is given by
\begin{align}
&S= \frac{1}{\hbar} \int_{\Sigma\times C} \omega \wedge \textrm{CS}(A)\;.
\end{align}

The theory is defined on a product manifold of the form $\Sigma \times C$,
where the theory is topological along $\Sigma$ and holomorphic along $C$.
The curve $C$ parametrizes the spectral parameter of the integrable model,
and is either $\mathbb{C}, \mathbb{C}^{\times}$ or 
an elliptic curve $E$ \cite{BelavinDrinfeld,Costello:2017dso},
for rational, trigonometric and elliptic integrable models.

The one-form $\omega$ is a nowhere-vanishing holomorphic one-form on the curve $C$,
and this is wedged with the Chern-Simons three-form $\textrm{CS}(A)$ 
\begin{align}
\textrm{CS}(A)=\textrm{Tr}\left( A\wedge \d A+\frac{2}{3}A\wedge A\wedge A \right) \;.
\end{align}
where $A$ is the gauge field in the adjoint representation of some gauge group $G$.

The two-dimensional integrable models are obtained on the curve $\Sigma$, which we can take to be $\mathbb{R}^2$. On this plane we make a statistical lattice by  straight
Wilson lines along, say, horizontal and vertical directions.
These Wilson lines are  
located at specific points on the curve $C$,
which can be identified with the spectral parameters associated with the lines.
The vacuum expectation values of the Wilson lines
gives the statistical partition functions of the associated integrable lattice models. 

The four-dimensional viewpoint provides a rather elegant conceptual explanation for
the integrability of the model---the topological invariance of the theory along the curve $\Sigma$,
together with the existence of the transverse direction along the spectral curve $C$, automatically ensures
integrability (Yang-Baxter equation) of the model.

The goal of this short note is to connect this four-dimensional theory
to the more standard three-dimensional Chern-Simons theory.

Our discussion is partly motivated by the old literature, 
where one obtained knot invariants from suitable limits of the integrable models (where the Yang-Baxter equation reduces to the braid group relation, see \eg \cite{Wadati:1989ud} for a review).\footnote{Indeed, many of the early-day literature on knots invariants are partly motivated by integrable models.}
Since knot invariants are described by the three-dimensional Chern-Simons theory, one might hope this limiting procedure can also be interpreted in the field theory. In other words,
one might hope that a suitable modification of the three-dimensional Chern-Simons theory, but still in three dimensions, will be sufficient for understanding key properties of integrable models.\footnote{While the three-dimensional Chern-Simons theory has mostly successfully applied to knot theory \cite{Witten:1988hf},
there has been some attempts in the past to discuss integrable models inside the three-dimensional framework (see \eg \cite{Witten:1989wf,Witten:1989rw}).}

Our paper relies on the following simple statement:
the four-dimensional Chern-Simons theory \eqref{4d_CS} is
``T-dual'' to an analytic continuation of the more standard three-dimensional Chern-Simons theory
\begin{align}
&S_{\rm 3d}= \frac{1}{\hbar_{\rm 3d}} \int_{\mathbb{R}^3 }  \textrm{CS}(A)\;.
\label{3d_CS}
\end{align}
where $\hbar_{\rm 3d}$ is the inverse level of the Chern-Simons theory.

In the rest of this paper we discuss this T-duality first in
field theory (section \ref{sec.FT}), and then in string theory (section \ref{sec.TS}).
This gives string theory realization of the four-dimensional theory \eqref{4d_CS}.
We also comment on the T-duality for more general theories in section \ref{sec.general}.

\section{T-duality in Field Theory}\label{sec.FT}

In this section we discuss this T-duality at the field theory level. 

For concreteness 
in this section we concentrate on the case where the spectral curve is $C=\mathbb{C}^{\times}\simeq \mathbb{R}\times S^1\simeq T^* S^1$.
The integrable models are then trigonometric, and are associated with the 
symmetries of the infinite-dimensional quantum affine algebras $U_q^{\rm  aff}(\mathfrak{g})$ (here $\mathfrak{g}$ is the 
Lie algebra of the Gauge group $G$), and $q$ is the quantum deformation parameter for the algebra.
Since this contains in particular
the quantum group $U_q(\mathfrak{g})$ as a subgroup, and since
the quantum group $U_q(\mathfrak{g})$ (at root of unity) is related to the Chern-Simons theory \cite{Witten:1989rw,ReshetikhinTuraev}, 
this would be the ideal starting point for our exploration.

The holomorphic one-form is given simply by $\omega=dz/z$ in a holomorphic coordinates on $\mathbb{C}^{\times}$, 
and we have 
\begin{align}
&S_{\rm 4d}= \frac{1}{\hbar} \int_{\mathbb{R}^2\times \mathbb{C}^{\times}}  \frac{\d z}{z} \wedge \textrm{CS}(A)\;.
\label{4d_CS}
\end{align}

\subsection{T-duality Prescription}

Let first start from the three-dimensional Chern-Simons theory
with gauge group $G=U(N)$, defined on $\mathbb{R}^3$ with 
coordinates $x, y, t$. Written in components, we have
\begin{align}
S_{\rm 3d}
&=\frac{1}{\hbar_{\rm 3d}} \int_{\mathbb{R}^3} \textrm{Tr}\left(
A_x \partial_t A_y +A_t(\partial_x A_y -\partial_y A_x)+[A_x, A_y] A_t 
\right) \;.
\end{align}
We will T-dualize along one of the transverse directions (which we call $\theta$), to obtain  a
four-dimensional theory on a four-dimensional spacetime on $\mathbb{R}^3\times S^1$ with coordinates $x, y, t, \theta$.
We will denote the size of the $S^1_{\theta}$ by $R'$.

Our T-duality prescription in field theory is motivated by the work of 
W.~Taylor, who considered a T-duality 
for Yang-Mills Higgs theory \cite{Taylor:1996ik}. There are, however,
important differences as we shall discuss below.\footnote{Taylor's T-duality for 3d Chern-Simons theory was discussed for example in \cite{Ishii:2007sy}. This was however the T-duality between
2d BF theory and 3d Chern-Simons theory, where the situation is much closer to the T-duality for the Yang-Mills-Higgs system in \cite{Taylor:1996ik}. Our T-duality is between 3d Chern-Simons theory and 4d Chern-Simons theory, and is possible only at the cost of breaking the 3d covariance of the 3d Cher-Simons theory, as we will see momentarily.}

The prescription consists of two steps.

\noindent {\bf Step 1:}

First, we go to an infinite-dimensional covering space of $\mathbb{R}^3$, where each component $A_{i=x,y,t}$ of the 
gauge field is promoted to an infinite copy of gauge fields $A_i^{m, n}$ moving in the covering space,
with $m, n$ running over all integers.
This infinite copy arises since the circle $S^1_{\theta}$ (along which we take T-duality)
is $S^1=\mathbb{R}/ 2\pi \mathbb{Z}$, with $\mathbb{R}$ is the universal cover and 
the effect of the quotient by $2\pi \mathbb{Z}$ creates an infinite copy of mirror images.
The field $A_i^{m,n}$ represents the winding mode from the $m$-th copy into the $n$-th copy, 
with winding number $m-n$, and we now have a gauge theory with infinite rank gauge group, 
with $G$ placed at each sheet.

Each term in the Lagrangian will be promoted to a term in the covering space.
For example, the term $\textrm{Tr}\left(A_x A_y A_t \right)$ is promoted to 
$\textrm{Tr}\left( A^{m, p}_x A^{p, q}_y A^{q, m}_t \right)$, which represents interaction between
$m, p$ and $q$-th sheets (note that for gauge invariance we should come back to the same sheet
inside the trace, and hence for example $\textrm{Tr}\left( A^{m, p}_x A^{p, q}_y A^{q, n}_t \right)$ with $m\ne n$ is not allowed). Following this procedure for each term, we obtain an action
\begin{align}
S=
\frac{2}{\hbar_{\rm 3d}}
\int_{\mathbb{R}^3}\textrm{Tr}\left(
A^{m, n}_x \partial_t A^{n, m}_y +A^{m, n}_t(\partial_x A^{m, n}_y -\partial_y A^{m, n}_x)
+A^{m, p}_x A^{p, q}_y A^{q, m}_t
-  A^{m, p}_y A^{p, q}_x A^{q, m}_t
\right) \;.
\end{align}
Note that we have used the Einstein conventions, with summations over indices $m,n, \dots$ implicit.

\begin{figure}[htbp]
\centering\includegraphics[scale=0.65]{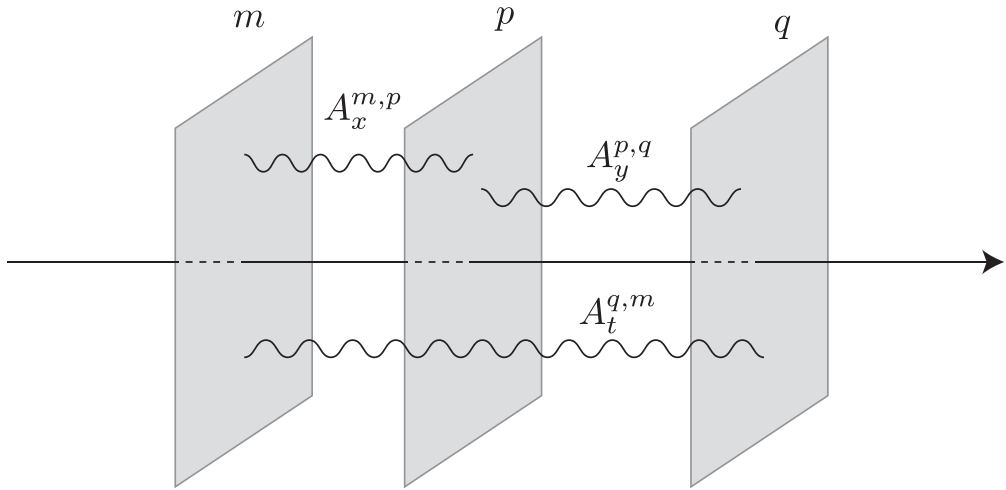}
\caption{In Step 1 of the T-duality, we go to the covering space, where each copy of the theory 
is labeled by an integer $m$, so that the original adjoint gauge field $A_{i=x,y,t}$ are now upgraded to
$A_{i=x,y,t}^{m,n}$ connecting $m$-th and $n$-th copies. We change the Lagrangian accordingly---for example, 
the term $\textrm{Tr}\left(A_x A_y A_t \right)$ in the Lagrangian is replaced by $\textrm{Tr}\left( A^{m, p}_x A^{p, q}_y A^{q, m}_t \right)$,
which describes the interaction between $p, q$ and $m$-th copies.}
\label{fig.winding}
\end{figure}

\bigskip

\noindent {\bf Step 2:}

In the second step, we impose the periodicity condition
\begin{align}
&A_{i}^{m,n}=A_{i}^{m-1,n-1}\quad (i=x,y) \;, \label{nowinding}\\
&A_{t}^{m,n}=A_{t}^{m-1,n-1} \quad (m\ne n) \;, \quad
A_t^{n,n}=(2\pi R') \, {\rm Id}_{N\times N}+ A_t^{n-1,n-1} \;. \label{winding}
\end{align}
Here the effect of the winding is represented by the condition for the diagonal 
component of $A_t$. We introduced a parameter $R'$, which plays the role of the 
radius of the extra dimensions.
Note that here the components $A_{x,y}$ and 
$A_t$ are treated asymmetrically here.

By using the periodicity conditions \eqref{nowinding} and \eqref{winding},
the Lagrangian is now written in terms of the fields $A_i^{m}:=A_i^{m, 0}$ only, 
and we obtain
\begin{align}
\label{from_3d}
S
=
\frac{2}{\hbar_{\rm 3d}}\int_{\mathbb{R}^3} \textrm{Tr}\left(
A^{m}_x (\partial_t-4 m \pi R')  A^{-m}_y +A^{m}_t(\partial_x A^{-m}_y -\partial_y A^{-m}_x)
+A^{m}_x A^{n}_y A^{-m-n}_t
-  A^{m}_y A^{n}_x A^{-m-n}_t 
\right) \;.
\end{align}

\paragraph{Comparison with 4d}

Let us now compare the theory we obtain \eqref{from_3d} with the four-dimensional theory
\eqref{4d_CS} on $\mathbb{R}^2\times \mathbb{C}^{\times}$.

We choose coordinates $x,y$ for the topological direction $\mathbb{R}^2$,
and the holomorphic coordinate of $\mathbb{C}^{\times}$ to be $z=e^{(t+\i \theta)/R}$,
where $t$ and $\theta$ are the cylindrical coordinates on $\mathbb{C}^{\times}\simeq \mathbb{R}\times S^1$
($\theta$ has period $2\pi R$).
Note that the radius of the cylinder is set by the scale $R$.

Let us write the action \eqref{4d_CS} in terms of the three components of the gauge connection
\begin{align}
A=A_x\d x+A_y \d y+A_{\bar{z}} \left( R \frac{\d\bar{z}}{\bar{z}} \right) \;.
\label{A_component}
\end{align}
Here we inserted a factor of $R$ to keep track of the dimensions.
We obtain (again we inserted a factor of $R$ in front of $\d z/z$, to keep track of dimensions)
\begin{align}
S&=\frac{2}{\hbar}\int_{\mathbb{R}^2 \times \mathbb{R}_t\times S^1} 
\left(\d x \wedge \d y\wedge R \frac{\d z}{z} \wedge R\frac{\d\bar{z}}{\bar{z}}\right)\,
 \textrm{Tr}\left(
A_{\bar{z}} (\partial_x A_y- \partial_y A_x) -A_x \bar{z}\partial_{\bar{z}} A_y +
[A_x, A_y] A_{\bar{z}} 
\right) \nonumber\\
&=\frac{4 \i}{\hbar}\int_{\mathbb{R}^2 \times \mathbb{R}_t\times S^1} 
\left(\d x \wedge \d y\wedge \d t \wedge \d\theta\right)\,
 \textrm{Tr}\left(
A_{\bar{z}} (\partial_x A_y- \partial_y A_x) -A_x \bar{z}\partial_{\bar{z}} A_y +
[A_x, A_y] A_{\bar{z}} \right) \;.
\end{align}

To perform T-duality, we expand the field in Fourier modes in the $\theta$-direction:
\begin{align}
A_x =\sum_{n \in \mathbb{Z}} A_x^n(x,y,t) e^{\frac{\i n \theta}{ R}} \;, \quad
A_y =\sum_{n \in \mathbb{Z}} A_y^n(x,y,t) e^{\frac{\i n \theta}{ R}} \;,\quad
A_{\bar{z}}=\frac{R}{2}\sum_{n \in \mathbb{Z}} A_t^n(x,y,t) e^{\frac{\i n \theta}{ R}}\;,
\end{align}
where in the Fourier expansion of $A_{\bar{z}}$ we included a factor of $R/2$ for later convenience. 

Using $\bar{z}\partial_{\bar{z}}=\frac{R}{2} \left(\frac{\partial}{\partial t} +\i \frac{\partial}{\partial \theta} \right)$ and integrating over $\theta$,
we obtain 
\begin{align}
\begin{split}
S&=\frac{4\pi \i  R }{ \hbar} \left[ \sum_{n\in \mathbb{Z}}  \int_{\mathbb{R}^2 \times \mathbb{R}_t} \textrm{Tr}\left(
A_x^{-n} \left(
\partial_t A_y^n -\frac{n}{R} A_y^n \right)
+ A_t^{-n} (\partial_x A_y^n- \partial_y A_x^n)
\right) \right.\\
&\qquad\qquad+ \left.
 \sum_{m, n\in \mathbb{Z}}
\int_{\mathbb{R}^2 \times \mathbb{R}_t} \textrm{Tr}\left(
[A_x^m, A_y^n] A_t^{-m-n}
\right)\right]\;.
\end{split}
\label{from_4d}
\end{align}

We can now compare the two results \eqref{from_3d}
and \eqref{from_4d}.
We then conclude that the two result coincides 
if we identify the two radii $R$ and $R'$ by 
\begin{align}
4\pi R \longleftrightarrow \frac{1}{R'} \;.
\label{TR}
\end{align}
and 
\begin{align}
\frac{\hbar}{2\pi \i  R} \longleftrightarrow \hbar_{\rm 3d} \;.
\label{hRk}
\end{align}
The inverse relation \eqref{TR} is indeed the T-duality relation one might have expected.
One can also identify the coupling constant $\hbar$ of the four-dimensional theory
with the inverse level $\hbar_{\rm 3d}$ of the three-dimensional theory.

We therefore conclude that the four-dimensional Chern-Simons theory \eqref{4d_CS}
is indeed T-dual to the three-dimensional Chern-Simons theory \eqref{3d_CS}.

\subsection{Comments on Subtleties}

Let us some important differences with the similar discussion of T-duality 
for the Yang-Mills-Higgs system \cite{Taylor:1996ik}.

In the original discussion of \cite{Taylor:1996ik}
considered the Yang-Mills-Higgs system, where
we have the adjoint scalars in addition to gauge fields.
In this case we can impose the winding mode condition \eqref{winding}
for some of the adjoint scalar fields, without spoiling the Lorentz symmetry of the theory.

This is in sharp contrast in our case, where the starting theory has only gauge fields (no matter fields).
We can still take one of the components of the adjoint gauge fields to 
play the role of the adjoint scalar, and this is what we did 
when imposing the winding condition in \eqref{nowinding} and \eqref{winding}.
However, this of course breaks the three-dimensional 
diffeomorphism symmetry along $x, y, t$: only two-dimensional diffeomorphism symmetry
along $x, y$ is preserved. This is, however, not a problem
for applications to integrable models \cite{Costello:2013zra,Witten:2016spx,Costello:2017dso,Costello:2018gyb}.

Another subtlety is that the action is complex after T-duality, as is evident from the fact that the 
four-dimensional action \eqref{4d_CS} treats the $z$ and $\br{z}$ components of the gauge field asymmetrically.
In other words, the three-dimensional theory is an analytic continuation of the compact group Chern-Simons theory, and we need to specify the choice of the integration contours in the space of complex gauge connections,
for a proper non-perturbative definition of the theory \cite{Witten:2010cx}.
This is related to the fact that the inverse level $\hbar_{\rm 3d}$ of the three-dimensional theory \eqref{3d_CS} is not quantized; see the relation \eqref{hRk}.

\subsection{Effective Three-Dimensional Theory}

From the three-dimensional perspective, the theory \eqref{from_4d}
contains the action for the zero-mode:
\begin{align}
\begin{split}
S^0&= \frac{4\pi \i R}{\hbar}\left[\sum_{n\in \mathbb{Z}}  \int_{\mathbb{R}^2 \times \mathbb{R}_t} \textrm{Tr}\left(
A_x^{0}\partial_t A_y^0 
+ A_t^{0} (\partial_x A_y^0- \partial_y A_x^0)
+
[A_x^0, A_y^0] A_t^{0}
\right)
\right]\;,
\end{split}
\end{align}
which is the three-dimensional Chern-Simons term for 
the three-dimensional connection $A^{0}(x,y,t):=A_{x}^0 \d x+A_{y}^0 \d y+A_{t}^0 \d t$, as expected.

We in addition have infinitely-many KK modes along the $S^1$, given by $A^m$ with $m\ne 0$.

When the radius $R$ is small, all these KK modes are very massive, and we expect that we can integrate out
these KK modes, to obtain an effective action for the zero modes (three-dimensional gauge fields).

In order to integrate out the non-zero modes, first note 
we can take advantage of the gauge symmetry to choose a gauge-fixing condition
$A_{t}^{n\ne 0}=0$ (Note that the zero-mode part $A_{t}^{0}$
is a holonomy of the four-dimensional gauge field along the $S^1_{\theta}$-direction, and hence is gauge invariant and cannot be gauged away.) Under this gauge, 
the non-zero-mode part of the action is
\begin{align}
\frac{S-S^0}{4\pi \i R/\hbar}&= \sum_{n\in \mathbb{Z}\backslash \{0 \}}  \int_{\mathbb{R}^2 \times \mathbb{R}_t} \textrm{Tr}\left(
A_x^{-n} \left(
\partial_t A_y^n -\frac{n}{R} A_y^n \right)
+ 
[A_x^{-n}, A_y^n] A_t^{0}
\right)\nonumber\\
&= \sum_{n\in \mathbb{Z}\backslash \{0 \}}  \int_{\mathbb{R}^2 \times \mathbb{R}_t} \textrm{Tr}\left(
A_x^{-n} \left(
\partial_t -\frac{n}{R} -\textrm{ad}( A_t^{0})\right)A_y^n
\right)\;,
\end{align}
where $\textrm{ad}$ denotes the adjoint action. Note that the zero-modes do not appear in this expression.

This action is quadratic in $A_{x,y}^{n\ne 0}$, and hence we can easily integrate out the non-zero modes, to obtain
\begin{align}
\frac{S-S^0}{4\pi \i  R/\hbar}&= \sum_{n\in \mathbb{Z}\backslash \{0 \}}  
-\log \textrm{det} \left(
\partial_t -\frac{n}{R} -\textrm{ad}( A_t^{0})\right) \nonumber\\
&\sim 
-\log \textrm{det} \sin\left( \pi R
D_t \right) 
\nonumber \\
&\sim 
-\textrm{tr} \log \left[ 1- e^{ -2\pi R D_t } \right] 
\;,
\label{eff_action}
\end{align}
with $D_t:=\partial_t  -\textrm{ad}( A_t^{0})$.

The correction to 3d Chern-Simons comes in as order $\mathcal{O}(e^{-R})$, and this expansion is good when $R$ is large, 
or equivalently when $R'$ is small.

The effective action \eqref{eff_action} makes manifest the fact that we are discussing a three-dimensional theory
with corrections from the standard Chern-Simons term. Such an effective action will in principle be sufficient for 
reproducing the results in integrable models (\eg R-matrix), at least order by order in power series expansion in $e^{-R}$. 

\section{T-duality in String Theory}\label{sec.TS}

There is a natural question concerning the explanation of integrable models 
from the four-dimensional Chern-Simons theory: can we embed the setup into string theory?

In this section we discuss the T-duality as in the previous section,
but now in the context of string theory.\footnote{The string theory embedding of the four-dimensional theory was discussed also recently in \cite{Costello:2018txb}.}

Let us begin with the three-dimensional Chern-Simons theory
defined on $\mathbb{R}^3$, with gauge group $SU(N)$.
Thanks to the work of Witten \cite{Witten:1992fb}, we know that this theory can
be regarded as a theory of $N$ D-branes wrapping the base $\mathbb{R}^3$ of
$T^*\mathbb{R}^3\simeq \mathbb{C}^3$. We choose the coordinates of the base to be $x, y, t$,
and the fiber to be $p_x, p_y, p_t$. 
The base $\mathbb{R}^3$ is 
obviously a Lagrangian submanifold inside $T^*\mathbb{R}^3$, with the symplectic form
given by 
\begin{align}
\omega=\d x \wedge \d p_x+\d y \wedge \d p_y+\d t \wedge \d p_t\;.
\label{omega}
\end{align}

We can also incorporate the Wilson lines (knots inside the base $\mathbb{R}^3$) into the Chern-Simons theory. 
According to Ooguri and Vafa \cite{Ooguri:1999bv}, such Wilson lines can be 
identified with another set of D-branes wrapping Lagrangian submanifolds,
such that the intersection of these D-branes with the original $N$ D-branes 
is gives the knots in the base. Let us denote this Lagrangian submanifold by $L_{\rm base}$.

For our discussion of integrable models the base manifold is a non-compact manifold
$\mathbb{R}^3$. 
which are in straight lines along $x$ or $y$-directions. 
Let us represent the knots as Lagrangian submanifolds $L_x, L_y$, each spreading in the $x, y$-directions:\footnote{Here we consider straight Wilson lines. General knots are not allowed in the four-dimensional theory \eqref{4d_CS}, due to the framing anomaly \cite{Costello:2017dso}.}
\begin{align}
\begin{array}{c||ccc|ccc}
 & x & y & t & p_x & p_y & p_t \\
\hline 
\hline
L_{\rm base} &-  & - & - & & &  \\
\hline L_x &- & &  &  &- &-\\
\hline
L_y && - &  & - & &-\\
\hline
L_t &-& - &  &  & &- \\
\end{array}
\end{align}
We have also included a surface-like defect $L_t$ inside Chern-Simons theory.
We can again easily check that the submanifolds $L_x, L_y$ and $L_t$ are 
Lagrangian submanifolds with respect to the symplectic form \eqref{omega}.

Since the topological string theory is part of the full string theory \cite{Bershadsky:1993cx,Antoniadis:1993ze},
we can also embed the setup above into the string theory, following \cite{Ooguri:1999bv}.
We then have type IIA string theory on $T^*\mathbb{R}^3\times \mathbb{R}^4$,
where $T^* \mathbb{R}^3$ is the Calabi-Yau three-fold direction and $\mathbb{R}^4$ is the
transverse direction realizing four-dimensional $\mathcal{N}=4$ supersymmetry.
The D-branes wrapping Lagrangian submanifolds are 
now given by (1) for $x$ and $y$-directions, D4-branes filling $\mathbb{R}^2$ inside $\mathbb{R}^4$,
namely surface defects inside four-dimensional $\mathcal{N}=4$ theory,
and (2) for $z$-direction, D2-branes located at a point inside $\mathbb{R}^4$,
namely a local operator for the four-dimensional $\mathcal{N}=4$ theory:
\begin{align}
\begin{array}{c||cc|c|cc|c||cccc}
 & x & y & t & p_x & p_y & p_t  & x_0& x_1& x_2&x_3\\
\hline 
\hline
{\rm D4}_{\rm base} &-  & - & - & & &  & -& -& &\\
\hline {\rm D4}_x &- & &  &  &- &- &- &- & &\\
\hline
{\rm D4}_y && - &  & - & &- &- &- & &\\
\hline
{\rm D2}_t &-& - &  &  & &- & & & &\\
\end{array}
\end{align}

Let us T-dualize the setup.
As in the previous section, we T-dualize along the one of the transverse directions $\theta$.
This $\theta$ will then combine with one of the original coordinates, say $t$, into a complex combination.

Geometrically it is clear what should happen---since we are T-dualizing,
the coordinates canonically conjugate with $t$, namely $p_t$, are compactified.
This means we should choose $\theta=p_t$.

In the language of D-branes we can T-dualize along the $p_t$ direction.
We obtain the D5--D3 system:
\begin{align}
\begin{array}{c||cc|c|cc|c|cccc}
 & x & y & t & p_x & p_y & p_t=\theta &x_0 &x_1 &x_2 &x_3\\
\hline 
\hline
{\rm D5}_{\rm base} &-  & - & - & & &  -& -& -& &\\
\hline {\rm D3}_x &- & &  &  &- & &- &- & &\\
\hline
{\rm D3}_y && - &  & - & & &- &- & &\\
\hline
{\rm D1}_t &-& - &  &  & & & & & &\\
\end{array}
\label{branes}
\end{align}

In the language of topological string theory, our T-duality prescription makes the $(t, \theta)$ direction 
holomorphic (\ie B-model), while the remaining direction $(x,y, p_x, p_y)$ are still A-model.
We therefore obtained a mixture of the  topological A-model and B-model.

We propose that the brane configuration in  \eqref{branes} realizes the four-dimensional Chern-Simons-like theory \eqref{4d_CS}, together with the Wilson lines.
Indeed, restricted to the $(x,y, t, \theta)$-directions,
we obtain
the four-dimensional theory in $\mathbb{R}^2\times (\mathbb{R}\times S^1)$,
together with Wilson lines $W_{x, y}$ in $x, y$ directions
and a surface defect $S_t$ filling $\mathbb{R}^2$:
This is indeed the setup needed for the explanations of integrable models:
\begin{align}
\begin{array}{c||cc|cc}
 & x & y & t &  \theta \\
\hline 
\hline
\textrm{4d CS} &-  & - & - & - \\
\hline W_x &- & &  &  \\
\hline
W_y && - &  & \\
\hline
S_t &-& - &  & \\
\end{array}
\end{align}
In the discussion of the integrable lattice models in \cite{Costello:2017dso,Costello:2018gyb}
the two-dimensional statistical lattice is created out of the Wilson lines $W_x, W_y$ 
in the $x, y$-directions. The surface defect $S_t$ fills the two-dimensional directions, and 
creates two-dimensional integrable field theory, as will be discussed in detail in \cite{PartIII,PartIV} (see also \cite{StringsTalk}).

\section{More Examples of T-dualities}\label{sec.general}

While we phrased the discussion above as the T-duality between three dimensions
\eqref{3d_CS} and four dimension \eqref{4d_CS}, our discussion is much more general.

For example, we can T-dualize further to obtain a five-dimensional theory
\begin{align}
&S= \frac{1}{\hbar} \int_{\mathbb{R}\times \mathbb{C}^{\times}\times \mathbb{C}^{\times}} \frac{\d z_1}{z_1}\wedge \frac{\d z_2}{z_2} \wedge \textrm{CS}(A)\;,
\label{5d_CS}
\end{align}
where $z_1$ and $z_2$ are two holomorphic coordinates for the two $\mathbb{C}^{\times}$,
and $\mathbb{R}$ is the remaining topological direction. The five-dimensional gauge field $A$
still have three components: $A=A_x \d x +A_{\br{z}_1} \d\br{z}_1 +A_{\br{z}_2} \d\br{z}_2$,
since the $A_{z_1}$ and $A_{z_2}$ components drops out from the action \eqref{5d_CS}.
Such a five-dimensional theory was discussed by Costello \cite{Costello:2016nkh}.

Yet further T-duality generates a six-dimensional theory
\begin{align}
&S= \frac{1}{ \hbar} \int_{\mathbb{C}^{\times} \times \mathbb{C}^{\times}\times \mathbb{C}^{\times}} \frac{\d z_1}{z_1}\wedge \frac{\d z_2}{z_2} \wedge \frac{\d z_3}{z_3}\wedge  \textrm{CS}(A)\;,
\end{align}
where the gauge field is now completely holomorphic, $A=A_{\br{z}_1} \d\br{z}_1 +A_{\br{z}_1} \d\br{z}_2+A_{\br{z}_3} \d\br{z}_3$,
where $z_1, z_2, z_3$ are holomorphic coordinates on the three $\mathbb{C}^{\times}$'s.
This is nothing but the holomorphic Chern-Simons theory, which arises from D-branes in the topological B-model \cite{Witten:1992fb}.

In this sequence of three, four, five and six-dimensional theories, the we start with the topological Chern-Simons theory (A-model), and we gradually make the theory more and more holomorphic, until
we get a completely holomorphic theory (B-model) in six dimensions.\footnote{In this respect,
one might be tempted to call the four-dimensional theory
as an ``(AAB/3)-model'' and the five-dimensional theory by an ``(ABB/3)-model''.}

Instead of starting with the action \eqref{3d_CS}, one can start with other theories.
For example, one can start with the five-dimensional Chern-Simons action
\begin{align}
S_{\rm 5d}=\frac{1}{\hbar} \int_{\mathbb{R}^5} {\rm CS}_5 (A) \;,
\end{align}
where ${\rm CS}_5 (A)$ is the Chern-Simons five-form satisfying the descent relation
$\d {\rm CS}_5 (A)=\textrm{Tr} (F\wedge F\wedge F)$.
This is a topological theory along the whole $\mathbb{R}^5$.

After T-duality, we obtain a six-dimensional theory
\begin{align}
S_{\rm 5d}=\frac{1}{\hbar} \int_{\mathbb{R}^4\times \mathbb{C}^{\times}}  \frac{\d z}{z} \wedge {\rm CS}_5 (A) \;,
\end{align}
where the theory is now topological along $\mathbb{R}^4$ and holomorphic along $\mathbb{C}^{\times}$
(whose holomorphic coordinate we denoted by $z$).
We can again continue to T-dualize further, to obtain a higher-dimensional theory.

Similarly, we can start with the two-dimensional topological BF theory and obtain a three-dimensional theory 
after the T-duality.

As these discussions show, we expect that our T-duality prescription will work for a rather broad class of theories, mapping topological quantum field theories to mixed topological/holomorphic theory in higher dimensions. It would be an exciting topic for further exploration.

\section*{Acknowledgments}
MY would like to thank C.~Vafa for enlightening discussion,
whose observation motivated this note.
He would also like to thank K.~Costello and E.~Witten for collaboration in \cite{Costello:2017dso,Costello:2018gyb} and W.~Taylor  and  J.~Yagi for related discussion. He also thanks for several friends for encouragements to publish this note, which otherwise would have been never published.
The bulk of this project was completed during MY's stay at Harvard University during 2016 Fall---2017 Spring term, and he would like to thank Harvard University for hospitality, and ``JSPS Program for Advancing Strategic International Networks to Accelerate the Circulation of Talented Researchers'' for providing travel support. 
The research of MY was also supported by WPI program (MEXT, Japan), by JSPS KAKENHI Grant No.\ 15K17634 and 17KK0087, 19K03820, 19H00689 and by JSPS-NRF Joint Research Project.

\parskip=-2pt

\bibliographystyle{nb}
\bibliography{T_dual}

\end{document}